\documentclass[12pt,showpacs,aps,prc,onecolumn,nofootinbib]{revtex4-2}

\usepackage{graphicx} % Required for inserting images
\usepackage{amsmath}
\usepackage{dcolumn}% Align table columns on decimal point
\usepackage{bm}% bold math
\usepackage{mathtools}
\usepackage{url}

\newcommand{\m}[1]{\langle #1\rangle}
\newcommand{\intm}{\int_{-\infty}^{+\infty}}

\begin{document}
%\linenumbers
\title{Fuzzy logic for reconstructing arbitrary moments of multiplicity distributions}

\author{Anar Rustamov}
\email{a.rustamov@gsi.de, a.rustamov@cern.ch}
\affiliation{GSI Helmholtzzentrum f\"ur Schwerionenforschung, 64291
Darmstadt, Germany}

\begin{abstract}
The Identity Method is a statistical technique developed to reconstruct moments of multiplicity distributions of particles produced in high-energy nuclear collisions. The method leverages principles from fuzzy logic, allowing for a more nuanced representation of particle identification by assigning degrees of membership to different particle types based on detector signals. In this contribution, a new framework, based on a multivariate moment generation function, is developed that allows the derivation of the formulas used in the Identity Method in a more robust way. Moreover, within the introduced framework, the Identity Method is easily extended to cope with arbitrarily higher-order moments. The techniques developed here offer significant potential for improving the accuracy of multiplicity distribution analyses in high-energy nuclear collisions. While the primary focus of the work presented is on applications in high-energy particle and nuclear physics, it can also be applied in other areas where signal identification is probabilistic and data are noisy, such as in medical imaging, remote sensing, and various other fields of experimental science.

\end{abstract}

\maketitle

\section{Introduction}
In many areas of physics and beyond, the study of the moments or cumulants of joint probability distributions provides deep insight into the dynamics of the system under study. For example, the cumulants of multiplicity distributions of different particle species produced in nuclear collisions, such as those at the Large Hadron Collider at CERN, the Heavy Ion Synchrotron at GSI, or the Relativistic Heavy Ion Collider at BNL, are key observables used to study the dynamics and properties of quantum chromodynamic matter~\cite{Gross:2022hyw, Braun-Munzinger:2016yjz, Andronic:2017pug}. These fluctuations provide insights into the thermal and dynamical processes that occur in the hot, dense medium created in head-on collisions of heavy nuclei, including critical phenomena induced by phase transitions~\cite{Rustamov:2020ekv}. A key challenge in such studies is that the measured multiplicity distributions in experiments are typically biased due to incomplete information caused by the finite resolution of the measurement process. This leads to incomplete particle identification, where the signals obtained (such as energy deposition, particle velocity, etc.) cannot be unambiguously associated with a specific particle type, thus introducing uncertainties in the measurement of particle multiplicities and leading to biases in their probability distribution. There are two different ways to circumvent such difficulties, which include:

\begin{itemize}
    \item {\textbf {\textit {Refining track selection criteria:}}} To improve the purity and efficiency of reconstructed particles, traditional cut-based methods must be carefully optimized to avoid introducing biases that distort the ultimately measured cumulants. A common approach is to involve signals from additional detectors (such as time-of-flight detectors), which enhances the purity of the identified particles~\cite{STAR:2020tga}. However, this comes at the cost of reduced efficiency, as the need to match tracks across different detectors introduces matching inefficiencies. This reduction in efficiency necessitates the application of large correction factors, since the efficiency correction procedures scale with the cumulant order~\cite{Nonaka:2017kko}. More significantly, small efficiencies may lead to a loss of the signals from the critical phenomena being searched for, which cannot be recovered with efficiency corrections. Achieving the right balance between purity and efficiency is therefore critical for cut-based analyses.
    \item {\textbf {\textit {Advanced analysis techniques:}}} Tools like the Identity Method are crucial for mitigating the effects of purity and efficiency in fluctuation analyis~\cite{Gazdzicki:2011xz, Gorenstein:2011hr, Rustamov:2012bx, Arslandok:2018pcu, Gazdzicki:2019rrq}. Unlike traditional cut-based methods, the Identity Method offers a more sophisticated approach that can correct for incomplete particle identification without being adversely affected by low purity. As a result, it is particularly well-suited for studying higher-order cumulants, where statistical precision and the ability to handle complex experimental conditions are paramount.
\end{itemize}

This paper provides a general formulation of the incomplete probability distributions by constructing, for the first time, their multivariate moment generating function. This approach not only simplifies the derivation of the necessary formulas for the Identity Method, but also extends and generalises it. 

It is important to emphasize that particle identification is not the only challenge in fluctuation analyses. In recent years, substantial progress has been made to address other challenges, with prominent ones being correlations induced by conservation laws~\cite{Braun-Munzinger:2018yru, Braun-Munzinger:2020jbk, Vovchenko:2021kxx, Braun-Munzinger:2023gsd, Vovchenko:2024pvk} and non-critical contributions arising from participant or volume fluctuations~\cite{Gorenstein:2011vq, Skokov:2012ds, Braun-Munzinger:2016yjz, Rustamov:2022sqm, Holzmann:2024wyd}.

The paper is organized as follows:
\begin{itemize}
\item Section~\ref{secIntro} introduces the Identity Method in a nutshell.
\item Section~\ref{secFramework} presents a  new framework for deriving formulas within the Identity Method for arbitrarily higher-order moments of multiplicity distributions. Explicit formulas are provided up to the fourth-order moments.
\item Section~\ref{secConsclusions} provides a summary of the work.
\item All necessary notations and further derivations are given in Appendices~\ref{anotations} and~\ref{acumvsmom}.

\end{itemize}

\section{The Identity Method in a Nutshell}
\label{secIntro}
Instead of relying on binary classifications, where each detected signal is unambiguously assigned to a particular particle type, the Identity Method assigns degrees of membership to different particle types based on the observed detector signals~\cite{Gazdzicki:2011xz, Gorenstein:2011hr, Rustamov:2012bx, Arslandok:2018pcu, Gazdzicki:2019rrq}. In this way, particles are counted using floating-point numbers instead of integers. This is achieved by introducing proxies for particle multiplicities in a given event, the so-called 
$W$ quantities~\cite{Gorenstein:2011hr, Rustamov:2012bx}. For a given event, the $W$ quantity for a particle type $j$ is defined as:

\begin{equation}
\label{id1}
    W_{j} = \sum_{i=1}^n\omega_{j}(x_{i})\;,
\end{equation}\\
where the summation runs over all $n$ tracks in the given event, $x_{i}$ denotes the measured signal for a given track $i$ used for particle identification.  The variables $\omega_{j}\in[0,1]$ represent the probability that a given measurement $x_{i}$ belongs to a particle type $j$:

\begin{equation}
\label{id2}
\begin{aligned}
   &\omega_{j}(x_{i})=\frac{\rho_{j}(x_{i})}{\sum_{j}\rho_{j}(x_{i})}\;, \\\\
   &\int_{-\infty}^{+\infty} \rho_{j}(x)dx = \m{N_{j}}\;,
\end{aligned}
\end{equation}\\
where $\m{N_{j}}$ is the mean number (first-order raw moment) of particle type $j$, $\rho_{j}(x)$ functions represent the inclusive distribution of the particle identification signal obtained by fitting the corresponding measurements in the experiments. The particular choice of $\rho(x)$ functions may differ from experiment to experiment. For example, in the NA49~\cite{Anticic:2013htn}, NA61/SHINE~\cite{NA61SHINE:2020cxu} and ALICE~\cite{ALICE:2019nbs} experiments, the energy loss of particles in the gas volume of time projection chambers is used as the primary particle identification signal.

Fuzzy logic is contained in Eq.~\ref{id2}. In fact, this equation implies that particles can be counted using floating-point numbers instead of the usual integers. It turns out that this aspect of allowing fuzzy logic in particle counting makes it possible to solve the misidentification problem for moments/cumulants exactly, i.e. analytically and uniquely. 

The $W_{j}$ quantities for each particle type $j$ can be measured experimentally, allowing their averages over the events to be calculated, i.e. moments of any order of the $W$ distributions can be measured. In addition, any moment of the joint $W$ distributions for different particles can also be reconstructed. For example, one can measure the second-order raw moments, such as $\langle W_j^2 \rangle$, and the third-order raw moments, $\langle W_j^3 \rangle$, or a second-order joint raw moments $\m{W_{j}W_{k}}$, and so on.

By confronting such experimentally measured moments of the $W$ distributions with those calculated analytically, one obtains the sought-after moments of the true multiplicity distributions. The main difficulty of the method is therefore the derivation of these analytic expressions, which relate the moments of the $W$ and multiplicity distributions via a set of linear equations. The analytic relations for the second-order moments were obtained in Ref.~\cite{Gorenstein:2011hr}, which was further developed to include the third- and higher-order moments~\cite{Rustamov:2012bx,Mackowiak-Pawlowska:2017dma}. In all these papers the multinomial theorem is used to extract analytical expressions which become computationally intensive for higher-order moments. 

In this paper the multivariate moment generating function for the $W$ quantities are derived for the first time. This approach allows one to obtain an analytic expression for arbitrarily higher-order moments of $W$ distributions by simply evaluating derivatives of the moment generating function of the appropriate order with respect to the introduced auxiliary variables.

The first application of the Identity Method was carried out by the NA49 collaboration at the CERN SPS~\cite{Rustamov:2013len, Anticic:2013htn} and later by the ALICE collaboration at the CERN LHC~\cite{ALICE:2017jsh} and the NA61/SHINE collaboration at the CERN SPS~\cite{NA61SHINE:2020cxu}. In subsequent years, the method was further developed to reconstruct the fluctuations of net-baryon number~\cite{Arslandok:2018pcu}, with its applications reported in Refs.~\cite{Rustamov:2017lio, ALICE:2019nbs, ALICE:2022xpf}.

\section{A framework for arbitrarily higher-order moments}
\label{secFramework}

In this section, a new mathematical formalism is introduced that allows one to quickly derive the moments of the $W$ distributions for any number of particles and arbitrarily high-order moments. For the sake of clarity, we will first assume that we have only two types of particles in an event, such as $N_{\pi}$ pions and $N_{p}$ protons. The quantities $W_{p}$ and $W_{\pi}$, defined in Eq.~\ref{id1}, are then calculated as follows:

\begin{equation}
\begin{aligned}
    &W_{p} = \sum_{i=1}^{N_{p}}\omega_{p}(x_{i}^{p})+ \sum_{i=1}^{N_{\pi}}\omega_{p}(x_{i}^{\pi}), \\\\
    &W_{\pi} = \sum_{i=1}^{N_{p}}\omega_{\pi}(x_{i}^{p})+ \sum_{i=1}^{N_{\pi}}\omega_{\pi}(x_{i}^{\pi})\;,
    \end{aligned}
\end{equation}\\
where ($x_{1}^{p}, x_{2}^{p},...,x_{N_{P}}^{p}$) and ($x_{1}^{\pi}, x_{2}^{\pi},...,x_{N_{\pi}}^{\pi}$) are actual measurements per event of the particle identification signal for protons and pions, respectively. Thus, the joint moment generating function $M\left(t_{p},t_{\pi}\right)$ can be expressed as:
\begin{equation}
\label{momgenPrPi}
\begin{aligned}    &M\left(t_{p},t_{\pi}\right)=\sum_{N_{p},N_{\pi}=0}^{\infty}P(N_{p},N_{\pi})\prod_{i=1}^{N_{p}}\intm dx_{i}^p\mathcal{P}_{p}(x_{i}^{p})\times\prod_{j=1}^{N_{\pi}}\intm dx_{j}^\pi\mathcal{P}_{\pi}(x_{j}^{\pi}) \\ \\
&\times e^{t_{p}\left[\sum_{i=1}^{N_{p}}\omega_{p}(x_{i}^{p})+ \sum_{i=1}^{N_{\pi}}\omega_{p}(x_{i}^{\pi})\right]}\times e^{t_{\pi}\left[\sum_{i=1}^{N_{p}}\omega_{\pi}(x_{i}^{p})+ \sum_{i=1}^{N_{\pi}}\omega_{\pi}(x_{i}^{\pi})\right]} \\ \\
&=\sum_{N_{p},N_{\pi}=0}^{\infty}P(N_{p},N_{\pi})\left[\intm e^{t_{p}\omega_{p}(x) + t_{\pi}\omega_{\pi}(x)}\mathcal{P}_{p}(x)dx\right]^{N_{p}} \\\\
&\times\left[\intm e^{t_{p}\omega_{p}(x) + t_{\pi}\omega_{\pi}(x)}\mathcal{P}_{\pi}(x)dx\right]^{N_{\pi}}\;,\\
    &
\end{aligned}
\end{equation}
where $t_{p}$ and $t_{\pi}$ are auxiliary variables, $P(N_{p},N_{\pi})$ is the joint probability distribution of $N_{p}$ and $N_{\pi}$, while $\mathcal{P}_{j}(x)$ denotes the probability density function of a PID variable $x$  for particle type $j$ and is defined as the normalised $\rho_{j}(x)$ function:

\begin{equation}
    \mathcal{P}_{j}(x) \equiv \frac{\rho_{j}(x)}{\langle N_{j}\rangle}\;.
\end{equation}

The multivariate moment generating function for the vector of $W$ quantities $\vec{W} = (W_{1}, W_{2}, ...,W_{n})$, where the $j^{th}$ components of $\vec{W}$ refer to the particle type $j$, is the extension of Eq. ~\ref{momgenPrPi} to an arbitrary number of particles:

\begin{equation}
\label{eqmultimoment}
    M_{W}(t_{1},t_{2},...,t_{n}) =\sum_{N_{1}, N_{2}, ..., N_{n}=0}^{\infty} P(N_{1},N_{2}, ..., N_{n})\prod_{i=1}^{n}\left[\intm e^{\sum_{j=1}^n t_{j}\omega_{j}(x)}\mathcal{P}_{i}(x)dx\right]^{N_{i}}\;,
\end{equation}
where $P(N_{1},N_{2}, ..., N_{n})$ is a joint multiplicity distribution of $n$ particles, $t_{i}$s are auxiliary variables that facilitate the calculation of the moments of particle type $i$. To proceed further, the following functions are introduced:

\begin{equation}
\label{h_quantities}
    h_{i}(t_{1},t_{2},...,t_{n}) = ln\left[\intm e^{\sum_{j=1}^{n} t_{j}\omega_{j}(x)}\mathcal{P}_{i}(x)dx\right]\;.
\end{equation}

With the help of Eq.~\ref{h_quantities} the moment generating function of the multivariate $W$-distributions takes the form:

\begin{equation}
\label{momgenW_2}
\begin{aligned}
    &M_{W}(t_{1},t_{2},...,t_{n}) =  M_{N}[h_{1}(t_{1},t_{2},...,t_{n}),h_{2}(t_{1},t_{2},...,t_{n}),...,h_{n}(t_{1},t_{2},...,t_{n})] \\\\
    &=\sum_{N_{1}, N_{2}, ..., N_{n}=0}^{\infty} P(N_{1},N_{2}, ..., N_{n})\prod_{i=1}^{n} e^{h_{i}(t_{1},t_{2},...,t_{n})N_{i}}\;.
\end{aligned}
\end{equation}
 It is easy to see that the function $M_{N}[{h_{1}, h_{2},...,h_{n}}]$ is a moment generating function of the multiplicity distribution $P(N_{1},N_{2},. .,N_{n})$ with the corresponding auxiliary variables $h_{i}(t_{1}, t_{2},...,t_{n})$, while the latter are cumulant generating functions for $\omega$ variables. Thus, the $r^{th}$-order pure, $\m{N^{r}}$, and joint, $\m{N_{1}N_{2},...,N_{r}}$, raw moments\footnote{Raw moments are hereafter referred to as moments.} of the multiplicity distributions can be derived by taking the appropriate derivatives of the moment generating function $M_{N}[{h_{1}, h_{2},...,h_{n}}]$ with respect to the corresponding $h_{i}(t_{1}, t_{2},...,t_{n})$ functions evaluated at $h_{i}(t_{1}, t_{2},...,t_{n}) = 0$:

\begin{equation}
\label{momentsN}
    \begin{aligned}
      &\left<N_{i}^r\right> = \frac{\partial^r M_{N}(h_{i})}{\partial h_{i}^r}\bigg\rvert_{h_{i}=0}\;, \\ \\
&\left<N_{1}^{i_{1}}N_{2}^{i_{2}},...,N_{n}^{i_{n}}\right> = \frac{\partial^{(i_{1}+i_{2}+...+i_{n})} M_{N}(h_{1},h_{2},...,h_{n})}{\partial h_{1}^{i_{1}}\partial h_{2}^{i_{2}}...\partial h_{n}^{i_{n}}}\bigg\rvert_{\vec{h}=0}\;.
     \end{aligned}
\end{equation}

The cumulants of the $\omega$ variables are obtained in a similar way, by taking appropriate derivatives of the $h_{i}(t_{1}, t_{2},...,t_{n})$ functions with respect to the auxiliary variables $t_{j}$ and evaluating them at $t_{j} = 0$. Using the definitions of the $h_{i}$ functions from Eq.~\ref{h_quantities}, for example, for the first-order and second-order cumulants, we have:

\begin{equation}
\label{cumulantsOmega1}
\begin{aligned}
    &\kappa_{1}(\omega_{j;i}) = \frac{\partial h_{i}(t_{j})}{\partial t_{j}}\bigg\rvert_{t_{j}=0} =\intm\omega_{j}(x)\mathcal{P}_{i}(x)dx,\\ \\
    &\kappa_{2}(\omega_{j;i})= \frac{\partial^2 h_{i}(t_{j})}{\partial t_{j}^2}\bigg\rvert_{t_{j}=0} =  \intm\omega_{j}^2(x)\mathcal{P}_{i}(x)dx - \left(\intm\omega_{j}(x)\mathcal{P}_{i}(x)dx\right)^2,\\ \\
    &\kappa_{11}(\omega_{jk;i})= \frac{\partial^2 h_{i}(t_{j},t_{k})}{\partial t_{j}\partial t_{k}}\bigg\rvert_{t_{j}=t_{k}=0} = \intm\omega_{j}(x)\omega_{k}(x)\mathcal{P}_{i}(x)dx \\\\
    &-\intm\omega_{j}(x)\mathcal{P}_{i}(x)dx\intm\omega_{k}(x)\mathcal{P}_{i}(x)dx\;,
    \end{aligned}
\end{equation}
where $k_{n}({\omega_{j;i}})$ denote the $n^{th}$-order cumulant of the $\omega_{j}(x)$ variables, for a probability density function $\mathcal{P}_{i}(x)$, whereas $k_{11}({\omega_{jk;i}})$ denotes the joint second-order cumulant of the $\omega_{j}(x)$ and $\omega_{k}(x)$ variables, for a probability density function $\mathcal{P}_{i}(x)$. By introducing the corresponding raw moments $\m{\omega_{j;i}}$ and $\m{\omega_{jk;i}}$, Eq.~\ref{cumulantsOmega1}, can be written as:

\begin{equation}
\label{cumulantsOmega2}
\begin{aligned}
    &\kappa_{1}(\omega_{j;i}) = \m{\omega_{j;i}},\\ \\
    &\kappa_{2}(\omega_{j;i})= \m{\omega_{j;i}^2}-\m{\omega_{j;i}}^2,\\ \\
    &\kappa_{11}(\omega_{jk;i})= \m{\omega_{jk;i}} - \m{\omega_{j;i}}\m{\omega_{k;i}}\;.
    \end{aligned}
\end{equation}

In a similar manner, higher-order cumulants of the $\omega$ variables are obtained:

\begin{equation}
\label{cumulantsOmega}
    \begin{aligned}
     &\kappa_{r}(\omega_{j;i}) = \frac{\partial ^r h_{i}(t_{j})}{\partial t_{j}^{r}}\bigg\rvert_{t_{j}=0}, \\ \\
     &\kappa_{i_{1}i_{2}...i_{n}}(\omega_{N_{1}N_{2}...N_{n};i}) = \frac{\partial^{(i_{1}+i_{2}+...+i_{n})}h_{i}(t_{1},t_{2},..,t_{n})}{\partial t_{1}^{i_{1}}\partial t_{2}^{i_{2}}...\partial t_{n}^{i_{n}}}\bigg\rvert_{\vec{t}=0}\;,\\
     &
     \end{aligned}
\end{equation}

The relations between the cumulants and  moments of the $\omega$ quantities for higher orders can be obtained by using the standard relations between cumulants and moments, also given in Appendix~\ref{acumvsmom}.

In fact, Eq.~\ref{momgenW_2} establishes all necessary relations between the moments of the $W$ quantities and the moments of the multiplicity distributions that are being searched for. Some explicit examples are given in the following subsections.

\subsection{First-order moments}

The first-order moment of the $W$ distribution for particle type $a$, $\m{W_{a}}$, can be easily derived by taking the first derivative of the moment generating function with respect to the auxiliary variable $t_{a}$:

\begin{equation}
\label{first-orderDer}
     \frac{\partial M_{W}(t_{a})}{\partial t_{a}} = \frac{\partial M_{N}}{\partial h_{1}}\frac{\partial h_{1}}{\partial t_{a}} + \frac{\partial M_{N}}{\partial h_{2}}\frac{\partial h_{2}}{\partial t_{a}} + ... + \frac{\partial M_{N}}{\partial h_{n}}\frac{\partial h_{n}}{\partial t_{a}}= \sum_{i = 1}^n\frac{\partial M_{N}}{\partial h_{i}}\frac{\partial h_{i}}{\partial t_{a}}\;.
\end{equation}

Evaluating Eq.~\ref{first-orderDer} at $t_{a}=0, h_{i}=0$ and using Eqs.~\ref{momentsN} and,~\ref{cumulantsOmega1} gives:

\begin{equation}
\label{first_orderpure}
 \left<W_{a}\right> =\left <N_{1}\right>\kappa_{1}(\omega_{a;1}) + \left<N_{2}\right>\kappa_{1}(\omega_{a;2}) + ... + \left<N_{n}\right>\kappa_{1}(\omega_{a;n}) = \sum_{i=1}^n \left <N_{i}\right>\kappa_{1}(\omega_{a;i})\;,
\end{equation}
where $i$ runs over all particle species. The equation thus obtained is identical to Eq.~14 in Ref.\cite{Rustamov:2012bx}. The advantage here is that all higher-order moments of the $W$ distributions can be related to those of the multiplicity distributions in a similar way using the moment generating function. From Eq.~\ref{first_orderpure} it can be seen that:

\begin{equation}
\label{eq-equality}
    \m{W_{a}} = \sum_{i=1}^n \left <N_{i}\right>\kappa_{1}(\omega_{a;i}) =  \sum_{i=1}^{n}\intm\frac{\rho_{a}(x)\rho_{i}(x)}{\sum_{i=1}^n\rho_{i}(x)}=\intm \rho_{a}(x)dx = \m{N_{a}}\;,
\end{equation}
for higher-order moments, however, this equality does not hold.
 
\subsection{Second-order moments}

In this section, formulas are derived for the second-order moment of the $W$ quantity for particle type $a$ and the joint cumulant of the $W$ quantities for particle types $a$ and $b$. To calculate the second-order moment of $W_{a}$, $\m{W_{a}^2}$, we first need to find appropriate second order derivatives of the moment generating function:

\begin{equation}
\begin{aligned}
    &\frac{\partial^{2} M_{W}(t_{a})}{\partial t_{a}^2} = \sum_{i=1}^n\sum_{j=1}^n\frac{\partial^2 M_{N}}{\partial h_{j}\partial h_{i}}\frac{\partial h_{j}}{\partial t_{a}}\frac{\partial h_{i}}{\partial t_{a}} + \sum_{i=1}^n\frac{\partial M_{N}}{\partial h_{i}}\frac{\partial^2h_{i}}{\partial t_{a}^2}\;,
    \end{aligned}
\end{equation}
which, evaluated at $t_{a}=0, h_{i}=h_{j}=0$, allows us to calculate the second moment of $W_{a}$ using Eqs.~\ref{momentsN} and~\ref{cumulantsOmega1}:

\begin{equation}
\label{second-pure}
     \left<W_{a}^2\right>  = \sum_{i=1}^n\langle N_{i}^2\rangle\Omega_{i,i}^{a,a}+2\sum_{1\leq i<j\leq n} \langle N_{i}N_{j}\rangle\Omega_{i,j}^{a,a} + \sum_{i=1}^{n} \langle{N_{i}}\rangle\kappa_{2}(\omega_{a;,i})\;,
\end{equation}
where $\Omega_{i,j}^{a,b}$ is defined as:

\begin{equation}
\begin{aligned}
    &\Omega_{i,j}^{a,b}=\kappa_{1}(\omega_{a;i})\kappa_{1}(\omega_{b;j})\;.
\end{aligned}
\end{equation}

Now let's show an example of calculating joint cumulants between two different distributions of $W$, corresponding to particle species $a$ and $b$:

\begin{equation}
\begin{aligned}
    &\frac{\partial^2 M_{W}(t_{a},t_{b})}{\partial t_{b}\partial t_{a}} = \sum_{i=1}^n\sum_{j=1}^n\frac{\partial^2 M_{N}}{\partial h_{j}\partial h_{i}}\frac{\partial h_{j}}{\partial t_{b}}\frac{\partial h_{i}}{\partial t_{a}} + \sum_{i=1}^n\frac{\partial M_{N}}{\partial h_{i}}\frac{\partial ^2h_{i}}{\partial t_{b}\partial t_{a}}\;,
    \end{aligned}
\end{equation}
leading to:
\begin{equation}
\label{second-mixed}
\begin{aligned}
    \left<W_{a}W_{b}\right> = \sum_{i=1}^n\langle N_{i}^2\rangle\Omega_{i,i}^{a,b} + 
    \sum_{1\leq i<j\leq n} \langle N_{i}N_{j}\rangle\Omega_{P[i,j]}^{a,b} + \sum_{i=1}^{n} \langle{N_{i}}\rangle\kappa_{11}(\omega_{ab;i})\;,
    \end{aligned}
\end{equation}

where $P[i,j] \equiv [i,j] + [j,i]$ refers to a linear permutation, so that

\begin{equation}
    \Omega_{P[i,j]}^{a,b} = \Omega_{i,j}^{a,b}+\Omega_{j,i}^{a,b}\;.
\end{equation}

In the case of only two particle types $a$ and $b$ the second order moments are obtained by solving a system of three linear equations for $\m{W_{a}^2}$, $\m{W_{b}^2}$ and $\m{W_{a}W_{b}}$, where the only unknowns are $\m{N_{a}^2}$,  $\m{N_{b}^2}$ and $\m{N_{a}N_{b}}$~\cite{Gorenstein:2011hr}:

\begin{equation}
\label{sec-moment-matrix}
\begin{pmatrix}
    \m{N_{a}^2}\\
    \m{N_{b}^2}\\
    \m{N_{a}N_{b}}
    \end{pmatrix} =
    \begin{pmatrix}
    \Omega^{a,a}_{a,a} & \Omega^{a,a}_{b,b} & 2\Omega^{a,a}_{a,b}\\
    \Omega^{b,b}_{a,a} & \Omega^{b,b}_{b,b} & 2\Omega^{b,b}_{a,b}\\
    \Omega^{a,b}_{a,a} & \Omega^{a,b}_{b,b} & \Omega^{a,b}_{P[a,b]}
    \end{pmatrix}^{-1}
    \begin{pmatrix}
    \m{W_{a}^2} -\sum_{i=a,b}\m{N_{i}}\kappa_{2}(\omega_{a;i})\\
    \m{W_{b}^2} -\sum_{i=a,b}\m{N_{i}}\kappa_{2}(\omega_{b;i})\\
    \m{W_{a}W_{b}} -\sum_{i=a,b}\m{N_{i}}\kappa_{11}(\omega_{ab;i})
    \end{pmatrix}\;,
\end{equation}
where, the first-order moments $\m{N_{i}}$ can be estimated using Eq.~\ref{eq-equality}. Eq.~\ref{sec-moment-matrix} has no solutions when the functions $\rho_{a}(x)$ and $\rho_{b}(x)$ overlap completely. Indeed, in this case the matrix to be inverted is singular. In practice, however, such completely overlapping $\rho_{j}(x)$ functions do not occur.  

It is obvious that for $a=b$ Eq.~\ref{second-mixed} converges to Eq.~\ref{second-pure}. Therefore, in fact, it is sufficient to derive the formula only for $\langle W_{a}W_{b}\rangle$. 
Moreover, we note that by introducing higher order cumulants of $\omega$ variables, $\kappa_{n}(\omega_{i;j})$ and $\kappa_{nm}(\omega_{ij;k})$, as well as notations like $\Omega_{i,j}^{a,b}$ and $P[i,j]$, Eqs. ~\ref{second-pure} and~\ref{second-mixed} look much simpler than the corresponding ones in Ref~\cite{Gorenstein:2011hr}. As will be shown in the following sections, this is particularly noticeable for higher order moments.

\subsection{Third-order moments}

Here, by analogy with Eq.~\ref{second-mixed}, the expression for the $\langle W_{a}W_{b}W_{c}\rangle$ is derived. 

\begin{equation}
\label{eqW3}
\begin{aligned}
    \left<W_{a}W_{b}W_{c}\right> \equiv &\frac{\partial^3M_{W}}{\partial t_{c}\partial t_{b}\partial t_{a}}\bigg\rvert_{t_{a}=t_{b}=t_{c}=0} =  \sum_{i=1}^n \langle N_{i}^3\rangle\Omega_{i,i,i}^{a,b,c}+ \sum_{1\leq i<j<k\leq n}\langle N_{i}N_{j}N_{k}\rangle\Omega_{P[i,j,k]}^{a,b,c}  \\ 
    & +\sum_{1\leq i<j\leq n}\langle N_{i}^2N_{j}\rangle\Omega_{P[i,i,j]}^{a,b,c} + \sum_{1\leq i<j\leq n}\langle N_{i}N_{j}^2\rangle\Omega_{P[i,j,j]}^{a,b,c}  \\ 
    &+\sum_{i=1}^{n}\langle N_{i}^2\rangle\Omega_{i,i}^{Q[(ab),c]} + \sum_{1\leq i<j\leq n}\langle N_{i}N_{j}\rangle\Omega_{P[i,j]}^{Q[(ab),c]} + \sum_{i=1}^{n}\langle N_{i}\rangle\Omega_{i}^{(abc)}\;.
    \end{aligned}
\end{equation}

For the notations used in Eq.~\ref{eqW3}, such as $\Omega_{i,j,k}^{a,b,c}$, $P[i,j,k]$, etc., see Appendix~\ref{anotations}. If all these definitions are taken into account, Eq.~\ref{eqW3} is identical to Eq.~8 in Ref.~\cite{Rustamov:2012bx}.

\begin{itemize}
    \item Special cases
    \begin{itemize}
        \item Set $a=b=c$ to obtain $\langle W_{a}^3\rangle = \langle W_{a}W_{a}W_{a}\rangle$.
        \item Set $a=b$ and leave $c$ different to obtain $\langle W_{a}^2W_{b}\rangle = \langle W_{a}W_{a}W_{b}\rangle$\;.
    \end{itemize}
\end{itemize}

Thus, similar to $\langle W_{a}W_{b}\rangle$, having a formula for $\langle W_{a}W_{b}W_{c}\rangle$ is sufficient because it captures all possible third-order moments, and the specific cases follow by setting particular values for the indices $a$, $b$ and $c$. This principle can be extended to arbitrary higher-order moments as well. 

\subsection{Fourth-order moments}

Here, for the first time, an explicit expression is derived for the fourth-order moment $\langle W_{a}W_{b}W_{c}W_{d}\rangle$:

\begin{equation}
\label{eqW4}
\begin{aligned}
\langle W_{a}W_{b}W_{c}W_{d}\rangle \equiv &\frac{\partial^4M_{W}}{\partial t_{d}\partial t_{c}\partial t_{b}\partial t_{a}}\bigg\rvert_{t_{a}=t_{b}=t_{c}=t_{d} = 0} =\sum_{i=1}^n\langle N_{i}^4\rangle\Omega_{i,i,i,i}^{a,b,c,d} \\ 
&+ \sum_{1\leq i<j<k<l\leq n}\langle N_{i}N_{j}N_{k}N_{l}\rangle\Omega_{P[i,j,k,l]}^{a,b,c,d}  \\  
&+\sum_{1\leq i<j<k\leq n}\langle N_{i}^2N_{j}N_{k}\rangle\Omega_{P[i,i,j,k]}^{a,b,c,d}+ \sum_{1\leq i<j<k\leq n}\langle N_{i}N_{j}^2N_{k}\rangle\Omega_{P[i,j,j,k]}^{a,b,c,d}  \\ 
&+\sum_{1\leq i<j<k\leq n}\langle N_{i}N_{j}N_{k}^2\rangle\Omega_{P[i,j,k,k]}^{a,b,c,d} + \sum_{1\leq i<j\leq n}\langle N_{i}^2N_{j}^2\rangle\Omega_{P[i,i,j,j]}^{a,b,c,d} \\ 
&+\sum_{1\leq i<j\leq n}\langle N_{i}^3N_{j}\rangle\Omega_{P[i,i,i,j]}^{a,b,c,d} + \sum_{1\leq i<j\leq n}\langle N_{i}N_{j}^3\rangle\Omega_{P[i,j,j,j]}^{a,b,c,d}  \\ 
&+\sum_{i=1}^n\m{N_{i}^3}\Omega_{i,i,i}^{Q[(ab),c,d]}+ \sum_{1\leq i<j<k\leq n}\langle N_{i}N_{j}N_{k}\rangle\Omega_{P[i,j,k]}^{Q[(ab),c,d]}  \\  
&+\sum_{1\leq i<j\leq n}\langle N_{i}^2N_{j}\rangle\Omega_{P[i,i,j]}^{Q[(ab),c,d]} + \sum_{1\leq i<j\leq n}\langle N_{i}N_{j}^{2}\rangle\Omega_{P[i,j,j]}^{Q[(ab),c,d]} \\ 
&+\sum_{i=1}^n\m{N_{i}^2}\left(\Omega_{i,i}^{Q[(ab),(cd)]} + \Omega_{i,i}^{Q[(abc),d]}\right)\\
&+\sum_{1\leq i<j\leq n}\langle N_{i}N_{j}\rangle\left(\Omega_{P[i,j]}^{Q[(ab),(cd)]} + \Omega_{P[i,j]}^{Q[(abc),d]}\right)+\sum_{i=1}^n\langle N_{i}\rangle\Omega_{i}^{(abcd)}\;.
\end{aligned}
\end{equation}

For the notations used in Eq.~\ref{eqW4}, such as $\Omega_{i,j,k,l}^{a,b,c,d}$, $P[i,j,k,l]$, etc., see Appendix~\ref{anotations}.

\subsection{Arbitrary moments}

The main achievement of this article is the derivation, for the first time, of the moment generating function for the $W$ quantities as given in Eq.~\ref{momgenW_2}. This equation establishes a relationship between the arbitrarily high-order moments of the distributions of the $W$ quantities and the corresponding, as yet unknown, multiplicity distributions. This is exactly what is needed to reconstruct the moments of the multiplicity distributions. By measuring the moments of the $W$ distributions, these equations can be used to calculate the moments of the multiplicity distributions~\cite{Rustamov:2012bx}. 
On the other hand, a close examination of Eqs.~\ref{second-mixed},~\ref{eqW3} and~\ref{eqW4} shows that, using only established patterns, one can write down the corresponding expressions for any higher-order moments of $W$ quantities. This avoids the tedious process of computing derivatives of the moment-generating function.

\section{Conclusion}
\label{secConsclusions}
In summary, the new mathematical framework presented extends the Identity Method by introducing a multivariate moment generating function for the $W$ quantities. By using derivatives of this generating function with respect to auxiliary variables, it becomes possible to compute moments of any order, including joint moments. This  approach improves the scalability and robustness of the Identity Method, in particular for its technical implementation in analysis packages.
The formulas derived up to the fourth-order moments show recognisable patterns, facilitating alternative extension of the technique to higher-order multiplicity distributions. The validity of the obtained formulas for the fourth-order moments has been verified using the simulation procedure reported in Ref.~\cite{Rustamov:2012bx}.
The approach reported opens up new avenues for exploring the properties of strongly interacting matter. In particular, it facilitates the study of higher-order cumulants of net charge distributions, including joint distributions of different net charges such as joint net-baryon and net-strangeness distributions. This is crucial for understanding fluctuations in heavy ion collisions and probing the QCD phase diagram. The framework is versatile and has potential applications beyond particle physics, especially in areas where detection processes complicate signal identification.

\section*{ACKNOWLEDGMENTS}
I would like to express my gratitude to Marek Gazdzicki, Mark Gorenstein, and Stanislaw Mrowczynski for our collaborative efforts in formulating and developing the original concepts behind the Identity Method. I also extend my thanks to Mesut Arslandok, Peter Braun-Munzinger, Marvin Nabroth, Johanna Stachel, Joachim Stroth, and Nu Xu for the fruitful discussions not only on the Identity Method but also on the complexities of fluctuation analyses in the quest to explore the phase structure of strongly interacting matter.
\appendix

\section{Notations}
\label{anotations}
This section gives all the shorthand notations used in Eqs.~\ref{eqW3} and~\ref{eqW4}. The products of the first-order cumulants of $\omega$ quantities are denoted by $\Omega_{i,j,...}^{a,b,...}$:

\begin{equation}
\begin{aligned}
    &\Omega_{i,j,k}^{a,b,c}=\kappa_{1}(\omega_{a;i})\kappa_{1}(\omega_{b;j})\kappa_{1}(\omega_{c;k})\;,\\ 
     &\Omega_{i,j,k,l}^{a,b,c,d}=\kappa_{1}(\omega_{a;i})\kappa_{1}(\omega_{b;j})\kappa_{1}(\omega_{c;k})\kappa_{1}(\omega_{d;l})\;.\\ \\
\end{aligned}
\end{equation}

If more than one of the upper indices are combined in parentheses, they refer to the joint cumulants of the $\omega$ quantities

\begin{equation}
\begin{aligned}
    &\Omega_{i,j}^{(ab),c}=\kappa_{11}(\omega_{ab;i})\kappa_{1} (\omega_{c;j})\;, \\ 
    &\Omega_{i,j}^{(abc),d}=\kappa_{111}(\omega_{abc;i})\kappa_{1} (\omega_{d;j})\;, \\ 
    &\Omega_{i,j,k}^{(ab),c,d}=\kappa_{11}(\omega_{ab;i})\kappa_{1} (\omega_{c;j})\kappa_{1} (\omega_{d;k})\;, \\ 
    &\Omega_{i,j}^{(ab),(cd)}=\kappa_{11}(\omega_{ab;i})\kappa_{11} (\omega_{cd;j})\;, \\ 
    &\Omega_{i}^{(abc)}=\kappa_{111}(\omega_{abc;i})\;, \\
    &\Omega_{i}^{(abcd)}=\kappa_{1111}(\omega_{abcd;i})\;. 
    &
\end{aligned}
\end{equation}

Linear permutations are denoted by $P[i_{i},i_{2},...,i_{n}]$:

\begin{equation}
\begin{aligned}
    & P[i,j] \equiv [i,j] + [j,i]\;, \\ 
    & P[i,j,k] \equiv [i,j,k] + [i,k,j] + [j,i,k] + [j,k,i]  + [k,i,j] + [k,j,i]\;.
\end{aligned}
\end{equation}
In a similar way permutations for more than three numbers can be computed. The total number of combinations in $P[i_{i},i_{2},...,i_{n}]$ is $n!$. If there are $k$ repeating indices, then the number of combinations is $n!/k!$. For example, $P[i,i,j]$ has only three combinations:

\begin{equation}
\begin{aligned}
    & P[i,i,j] \equiv [i,i,j] + [i,j,i] + [j,i,i]\;,
\end{aligned}
\end{equation}

The notations $Q[(ab),c]$, etc., where two or more upper indices are enclosed in parentheses are defined as:

\begin{equation}
\label{defQ}
\begin{aligned}
    & Q[(ab),c] \equiv [(ab),c] + [(ac),b] + [(bc),a]\;, \\ 
     & Q[(ab),c, d] \equiv [(ab),c,d] + [(ac),b,d] + [(ad),b,c] + [(bc),d,a] + [(bd),a,c] + [(cd),a,b]\;, \\ 
& Q[(abc),d] \equiv [(abc),d] + [(abd), c] +[(bcd),a] +[(cda),b]\;, \\ 
& Q[(ab),(cd)] \equiv [(ab),(cd)] + [(ac),(bd)] + [(ad),(bc)]\;.
\end{aligned}
\end{equation}

If $P[i,j,..]$ and/or $Q[(ab)c,...]$ appear in the indices of the $\Omega$ notations, then all combinations must be considered. For example:

\begin{equation}
\begin{aligned}
   &\Omega_{P[i,j]}^{a,b} = \Omega_{i,j}^{a,b}+\Omega_{j,i}^{a,b}\;, \\
   &\Omega_{P[i,j]}^{Q[(ab),c]} = \Omega_{i,j}^{(ab),c}+\Omega_{j,i}^{(ab),c} + 
   \Omega_{i,j}^{(ac),b}+\Omega_{j,i}^{(ac),b}+
   \Omega_{i,j}^{(bc),a}+\Omega_{j,i}^{(bc),a}\;. 
    \end{aligned}
\end{equation}

\section{Relations between moments and cumulants}
\label{acumvsmom}
The moments and cumulants of a random variable are two ways of describing its distribution, and there are specific relationships between them.  Moments provide an alternative measure of the probability distribution, while cumulants provide its description that is particularly useful because they have certain desirable properties, such as additivity for independent random variables. Cumulants and moments are related by their respective generating functions. For a random variable $X$ they are defined as

\begin{equation}
    K_{X}(t) = ln\left[M_{X}(t)\right]\;,\;\;\;\; M_{X}(t)=\langle e^{tX}\rangle\;,
\end{equation}
where $K_{X}(t)$ and $M_{X}(t)$ are cumulant and moment generating functions, respectively,  while $t$ is an auxiliary variable. Cumulants (moments) are obtained as derivatives of $K_{X}(t)$ ($M_{X}(t)$) evaluated a $t=0$:

\begin{equation}
    \kappa_{n}(X)=\frac{d^nK_{X}(t)}{dt^n}\bigg\rvert_{t=0},\;\;\;\; \langle X^n\rangle=\frac{d^nM_{X}(t)}{dt^n}\bigg\rvert_{t=0}\;.
\end{equation}
Since the cumulant generating function is the logarithm of the moment generating function, cumulants are related to moments by derivatives of this logarithmic function. For example, for the first two cumulants we get:

\begin{equation}
    \begin{aligned}
    &\kappa_{1}(X)=\left[\frac{1}{M_{X}(t)}\frac{dM_{X}(t)}{dt}\right]\bigg\rvert_{t=0}=\langle X\rangle\;, \\ \\
    &\kappa_{2}(X)=\left[\frac{1}{M_{X}(t)}\frac{d^2M_{X}(t)}{dt^{2}}-\left(\frac{1}{M_{X}(t)}\frac{dM_{X}(t)}{dt}\right)^2\right]\bigg\rvert_{t=0} = \langle X^{2}\rangle - \langle X\rangle^2\;, \\
    &
    \end{aligned}
\end{equation}
in which we have used a trivial condition $M_{X}(0) = 1$.

A more compact relationship between cumulants and moments can be established using the Fa{\`a} di Bruno formula and Bell polynomials~\cite{Braun-Munzinger:2020jbk, Friman:2022wuc}:

\begin{equation}
\label{cumBell}
    \kappa_{n}(X) = \sum_{k=1}^{n}(-1)^{(k-1)}{(k-1)}!B_{n,k}\left(\langle X\rangle,\langle X^2\rangle, ..., \langle X^{n-k+1}\rangle\right)\;,
\end{equation}\\
where $B_{n,k}$ is the partial Bell polynomial of index $(n, k)$. Using Eq.~\ref{cumBell}, explicit relations between cumulants and moments up to order four are given in Table I. These relations can be used to convert cumulants of $\omega$ quantities, $\kappa_{n}(\omega_{i;j})$ into the corresponding moments $\m{\omega_{i;j}}$, $\m{\omega_{i;j}^2}$ etc. (see Eq.~\ref{cumulantsOmega2}).

\begin{center}
\begin{table}[h]
\begin{tabular}{||c|c||}
 \hline
 cumulants  & relations to moments   \\  
 \hline
 \hline
 $\kappa_{1}(a)$ & $\langle a \rangle$  \\ 
 \hline
 $\kappa_{2}(a)$ & $- \langle a\rangle^2 + \langle a^2\rangle$ \\ 
 \hline
 $\kappa_{3}(a)$ & $2 \langle a\rangle^3  - 3\langle a\rangle\langle a^2\rangle + \langle a^3\rangle $  \\ 
 \hline
 $\kappa_{4}(a)$ & $-6\langle a\rangle^4 + 12 \langle a\rangle^2\langle a^2\rangle - 3\langle a^2\rangle^2 - 4\langle a\rangle\langle a^3\rangle + \langle a^4\rangle$  \\ 
 \hline
 \hline
\end{tabular}
\caption{Relations between the cumulants and the moments up to the fourth order.}
\end{table}
\end{center}

The multivariate cumulant generating function is an extension of the cumulant generating function to the multivariate case, where we are dealing with a vector of random variables $\vec{X}$ instead of a single random variable $X$. Let $\vec X = (X_{1},X_{2},...,X_{n})$ be a vector of random variables.  The multivariate cumulant generating function $K_{\vec{X}}(\vec{t})$ is obtained as the logarithm of the multivariate moment generating function $M_{\vec{X}}(\vec{t})$: 

\begin{equation}
    K_{\vec{X}}(t_{1}, t_{2}, ...,t_{n})=ln\left[M_{\vec{X}}(t_{1},t_{2},...,t_{n})\right]\;,\;\;\;\;\; M_{\vec{X}}(t_{1},t_{2},...,t_{n})=\langle e^{\sum_{i=1}^n t_{i}X_{i}}\rangle\;,
\end{equation}
with $t_{1}, t_{2},...,t_{n}$ being auxiliary variables.

The mixed cumulants are obtained by taking partial derivatives of $K_{\vec{X}}(\vec{t})$ with respect to components of $\vec{t}$ evaluated at $\vec{t}=0$.

\begin{equation}
  \kappa_{i_{1}i_{2}...i_{n}}(X_{1},X_{2},...,X_{n})= \frac{\partial^{(i_{1}+i_{2}+...+i_{n})}K_{\vec{X}}(t_{1},t_{2},...,t_{n})}{\partial t_{1}^{i_{1}}\partial t_{2}^{i_{2}}...\partial t_{n}^{i_{n}}}\bigg\rvert_{\vec{t}=0}\;.
\end{equation}

Similarly, the mixed raw moments are obtained as the corresponding derivatives of $M_{\vec{X}}(\vec{t})$.

\begin{equation}
  \langle X_{1}^{i_{1}}X_{2}^{i_{2}},...,X_{n}^{i_{n}}\rangle= \frac{\partial^{(i_{1}+i_{2}+...+i_{n})}M_{\vec{X}}(t_{1},t_{2},...,t_{n})}{\partial t_{1}^{i_{1}}\partial t_{2}^{i_{2}}...\partial t_{n}^{i_{n}}}\bigg\rvert_{\vec{t}=0}\;.
\end{equation}

For example, the joint cumulant of the second-order can be obtained as:

\begin{equation}
    \kappa_{11}(X_{i},X_{j})=\frac{\partial^2 K_{X_{i},X_{j}}(t_{i},t_{j})}{\partial t_{i}\partial t_{j}}\bigg\rvert_{t_{1}=t_{2}=0} = \langle X_{i}X_{j}\rangle - \langle X_{i}\rangle\langle X_{j}\rangle\;.
\end{equation}

The relations between joint cumulants and joint moments up to order four are given in Table II, where for simplicity the random variables are represented as $a,b,c,d$. These relations can be used to convert  the joint cumulants of the $\omega$ quantities into the corresponding joint moments (see Eq.~\ref{cumulantsOmega2}). The other combinations of joint cumulants can be obtained by index substitution. For example, to obtain $\kappa_{12}(ab)$, one must take $c=b$ in $\kappa_{111}(a,b,c)$:

\begin{equation}
    \kappa_{12}(a,b) = \kappa_{111}(a,b,c)\bigg\rvert_{c=b}=2\m{a}\m{b}^2 - 2\m{ab}\m{b} - \m{b^2}\m{a} +\m{ab^2}\;.
\end{equation}

\begin{center}
\begin{table}[hbt]
\begin{tabular}{||c|c||}
 \hline
 joint cumulants & relations to joint moments  \\ 
 \hline
 \hline
 $\kappa_{11}(a,b)$ & $ \m{ab} - \m{a}\m{b}$  \\ 
 \hline
 $\kappa_{111}(a,b,c)$ & $ 2\m{a}\m{b}\m{c} - \m{ab}\m{c} - \m{bc}\m{a} - \m{ac}\m{b}  +\m{abc} $  \\ 
 \hline
 $\kappa_{1111}(a,b,c,d)$ & $-6\m{a}\m{b}\m{c}\m{d} +2\m{ab}\m{c}\m{d} + 2\m{ac}\m{b}\m{d}+2\m{ad}\m{b}\m{c}+2\m{bc}\m{a}\m{d}$ \\
 &$+2\m{bd}\m{a}\m{c} +2\m{cd}\m{a}\m{b}  - \m{abc}\m{d} - \m{abd}\m{c}-\m{bcd}\m{a} - \m{cda}\m{b}$\\
 &$- \m{ab}\m{cd}-\m{ac}\m{bd} - \m{ad}\m{bc} + \m{abcd}$ \\
 \hline
 \hline
\end{tabular}
\caption{Relations between the joint cumulants and the joint moments up to the fourth order. These relations can be written in a compact way using the notations introduced in Eq.~\ref{defQ}.}
\end{table}
\end{center}

%\bibliography{paper}
%

\end{document}